\documentstyle[prl,twocolumn,aps]{revtex}

\begin{document}

\draft

\def\x{{\bf x}}
\def\y{{\bf y}}
\def\R{{\bf R}}

\title{Duality and zero-point length of spacetime}
\author{T. Padmanabhan\thanks{paddy@iucaa.ernet.in}}
\address{IUCAA, Post Bag 4, Ganeshkhind, Pune 411 007, India.}
\maketitle
\date{\today}

\begin{abstract}
The action for a relativistic free particle of mass $m$
receives a contribution $-mds$
from a path segment of infinitesimal length $ds$. Using this action 
in a path integral, one can obtain the Feynman propagator for a spinless 
particle of mass $m$. 
If one of the effects of quantizing gravity is to introduce a minimum length
scale $L_P$ in the spacetime, then one would expect the segments of paths 
with lengths less than $L_P$ to be suppressed in the path integral. 
Assuming that the path integral amplitude is invariant under the 
`duality' transformation $ds\to L_P^2/ds$, one can calculate the 
modified Feynman propagator. 
I show that this propagator is the same as the one obtained by assuming 
that: quantum effects of gravity leads to modification of the spacetime 
interval $(x-y)^2$ to $(x-y)^2+L_P^2$. 
This equivalence suggests a deep relationship between introducing a
`zero-point-length' to the spacetime and postulating invariance of path 
integral amplitudes under duality transformations.
\end{abstract}

>From the fundamental constants $G, \hbar $ and $c$, one can form a 
quantity with dimensions of length, $L_P \equiv (G\hbar/c^3)^{1/2}$, 
which is expected to play a vital role in the `ultimate' theory of 
quantum gravity. 
Simple thought experiments indicate that it is not possible to devise 
experimental procedures which will measure lengths with an accuracy 
greater than about ${\cal O}(L_P)$~\cite{paddy87}. 
This result suggests that one could think of Planck length as some 
kind of ``zero-point length'' of spacetime. 
In some simple models of quantum gravity, $L_P^2$ does arise as a mean 
square fluctuation to spacetime intervals, due to quantum fluctuations 
of the metric~\cite{paddy85}. 
In more sophisticated approaches, like models based on string theory 
or Ashtekar variables, similar results arise in one guise or the other 
(see {\it e.g.},~\cite{rovelli95,ashtekar92}).  

The existence of a fundamental length  implies that processes involving 
energies higher than Planck energies will be suppressed and the 
ultra-violet behavior of the theory will be improved. 
All sensible models for quantum gravity provide a mechanism for good 
ultra-violet behavior, essentially through the existence of a 
fundamental length scale. One direct consequence of such an improved 
behavior will be that the Feynman propagator (in momentum space) will 
acquire a damping factor for  energies larger than Planck energy.

If the ultimate theory of quantum gravity has a fundamental length scale 
built into it, then it seems worthwhile  to formulate quantum field theory,
using this principle as the starting point. This could, for example, help 
in understanding some of the effects of quantizing gravity on the matter 
fields. 
I will show in this {\it Letter} that such a procedure leads to some 
interesting results. 
I will be using a spin-zero (scalar) field as a prototype but the general 
ideas can be extended easily for more complicated situations.

To keep things well-defined and general, I will work in a $D-$dimensional
 Euclidean space.  Feynman propagator $G({\bf x},{\bf y})$ for a spin-zero, 
free particle of mass $m$ in $D-$dimension is
\begin{equation}
G_{{\rm conv}}(\x , \y) = \int {d^D {\bf p}\over (2\pi)^D} {e^{-i{\bf p} 
\cdot ({\bf x} - {\bf y})} \over \left( p^2 + m^2\right)}.\label{eqn:contori}
\end{equation}
This propagator---which arises in the standard formulation of quantum 
field theory---does not take into account the existence of any fundamental 
length in the spacetime. 
Let us ask how this propagation amplitude could be modified if there exists 
a fundamental zero-point length to the spacetime. 
This is best done using the path integral expression for the Feynman 
propagator  
\begin{equation}
G_{{\rm conv}}(\x , \y) = \sum_{{\rm paths}} 
\exp - m\, s  (\x,\y),\label{eqn:qq}
\end{equation}
where $s(\x,\y)$ is the length of any path connecting $\x$ and $\y$.
To give meaning to the path integral we shall first introduce a cubic 
lattice with a lattice spacing $\epsilon$ in the $D-$dimensional 
Euclidean space. 
The propagator in the latticized spacetime is given by 
\begin{equation}
{\cal G}_{{\rm conv}}({\bf R}, \epsilon) 
= \sum_{N=0}^\infty C(N, {\bf R}) 
\exp \left[ - \mu (\epsilon) \epsilon N\right],\label{eqn:latori}
\end{equation}
where $C(N,{\bf R})$ is the number of paths of length $N\epsilon$ 
connecting the origin to the lattice point $\R = (n_1, n_2, \cdots, n_D)$ 
which is a $D-$dimensional vector with integer components. 
(The physical scale corresponding to ${\bf R}$ is ${\x=\epsilon \R}$.)
The scaling factor $\mu(\epsilon)$ acts as the mass parameter on the 
lattice.  
The propagator for the continuum has to be obtained by 
multiplying~(\ref{eqn:latori}) by a suitable measure $\cal{M}(\epsilon)$ 
and taking the limit $\epsilon \to 0$. Both the measure $\cal{M}(\epsilon)$ 
and the mass parameter on the lattice $\mu(\epsilon)$ should be chosen 
so as to ensure the finiteness of the limit. 
This procedure is straight forward to carry out (see e. g.,~\cite{paddy94}) 
and one obtains the Feynman propagator given in equation~(\ref{eqn:contori}).

In the above procedure, the weightage given for a path of length $l$ is 
$\exp (-ml)$ which is a monotonically decreasing function of $l$. 
The existence of a fundamental length $L_P$ would suggest that paths with 
length $l \ll L_P$ should be suppressed in the path integral. 
This can, of course, be done in several different ways by arbitrarily 
modifying the expression in equation~(\ref{eqn:latori}). 
In order to make a specific choice I shall invoke the following 
``principle of duality''. 
I will postulate that the weightage given for a path  should be invariant 
under the transformation $l\to L_P^2/l$. Since the original path integral 
has the factor $\exp (-ml)$ we have to introduce the additional factor 
$\exp (-mL_P^2/l)$. We, therefore, modify equation~(\ref{eqn:latori}) to
\begin{equation} 
{\cal G} (\R, \epsilon) = \sum_{N=0}^\infty C(N, {\bf R}) 
\exp \left[ - \mu (\epsilon) \epsilon N 
- {\lambda (\epsilon) \over \epsilon N} \right],\label{eqn:latmod}
\end{equation} 
where $\lambda(\epsilon)$ is a lattice parameter which will play the role 
of $(m L_P^2)$ in the continuum limit.

I will take this to be the basic postulate arising from the ``correct'' 
theory  of quantum gravity. It may be noted that the `principle of duality'
invoked here is similar to that which arises in string theories (though
not identical). 
In fact we may think of equation~(\ref{eqn:latmod}) as the simplest 
realization of duality for a free particle; we have demanded that the
existence of a weightage factor $\exp (-ml)$ necessarily requires the 
existence of another factor $\exp (-mL_P^2/l)$. 
We shall now study the consequences of the modifications we have 
introduced. 
 
To evaluate this path integral on the lattice we begin by noticing that 
the generating function for $C(N,{\bf R})$ is given by~\cite{paddy94}
\begin{eqnarray}
F^N & \equiv & \sum_\R C(N; \R) e^{i{\bf k}\cdot \R}\nonumber\\ 
&=& \biggl(e^{ik_1} + e^{ik_2} + \cdots 
+ e^{ik_D}\nonumber\\
& &\qquad\qquad\quad 
+\, e^{-ik_1} + e^{-ik_2} + \cdots + e^{-ik_D}\biggl)^N.
\end{eqnarray}
Therefore we can write 
\begin{eqnarray}
\sum_\R e^{i{\bf k}\cdot \R} {\cal G}(\R, \epsilon) 
&=& \sum_{N=0}^\infty e^{-\mu\epsilon N - (\lambda / \epsilon N)} 
\sum_\R C(N,\R) e^{i{\bf k}\cdot \R}\nonumber\\ 
&=& \sum_{N=0}^\infty e^{-N (\mu \epsilon - \ln F) 
- (\lambda/\epsilon N)}.\label{eqn:qsumone}
\end{eqnarray}
Thus, our problem reduces to evaluating the sum of the form 
\begin{eqnarray}
S(a,b) &\equiv& \sum_{n=0}^{\infty} 
\exp\left( -a^2 n - {b^2\over n}\right)\nonumber\\
&=& \sum_{n=1}^{\infty} \exp\left(-a^2 n 
- {b^2\over n}\right).\label{eqn:qamb}
\end{eqnarray}
This expression can be evaluated by some algebraic tricks~\cite{paddy96} 
and the answer is 
\begin{eqnarray}
S(a,b) &=&\int_{0}^{\infty} {k dk\over 2b^2} {J_0(k) 
e^{-(a^2 + k^2/4b^2)}\over \left[1 
- e^{-(a^2 + k^2/4b^2)}\right]^2}\nonumber\\ 
&=& {1\over \left( 1-e^{-a^2}\right)} 
- \int_0^\infty dq {J_1(q) \over 
\left[1 - e^{-(a^2 + q^2/4b^2)}\right]},\label{eqn:qanss}
\end{eqnarray}
where $J_\nu (x)$ is the Bessel function of order $\nu$. 
The first form of the integral  shows that the expression is well defined 
while the second form has the advantage of separating out the 
$b$-independent part as the first term. 
(Note that the two summations in~(\ref{eqn:qamb}) will differ by unity if 
$b=0$; the results in~(\ref{eqn:qanss}) will go over to the second summation 
in~(\ref{eqn:qamb}) if the limit $b\to 0$ is taken.)
Using the second form in equation~(\ref{eqn:qanss}) and introducing the 
continuum variables ${\bf x} = \epsilon {\bf R}$, ${\bf p} = \epsilon^{-1} 
{\bf k}$, we can write the propagator as the sum of two terms ${\cal G}
= {\cal G}_0 + {\cal G}_c$,  where
\begin{eqnarray}
{\cal G}_0 (\R) &=& \int {d^D {\bf k} \over (2\pi)^D} \left\{ {e^{-i{\bf k}
\cdot \R}\over 1 - 2 e^{-\mu \epsilon} \sum_{i=1}^D \cos k_i}\right\},\\
{\cal G}_c(\R) &=& - \int_0^\infty dq J_1(q) \int {d^D {\bf k} \over 
(2\pi)^D}\;  e^{-i{\bf k}\cdot \R}\nonumber\\ 
& &\quad\;\; 
\times\,\left\lbrace{1 \over {1 - 2 e^{-\epsilon(\mu +(q^2/4\lambda))} 
\sum_{i=1}^D \cos k_i}}\right\rbrace.
\end{eqnarray}
We now have to take the $\epsilon \to 0$ limit. 
The propagator ${\cal G}_0$ becomes, in the limit of small $\epsilon$
\begin{equation}
{\cal G}_0(\x) = \int { d^D {\bf p} \over (2\pi)^D} {A_1(\epsilon) 
e^{-i{\bf p}\cdot \x}\over p^2 + B_1(\epsilon)},
\end{equation}
where
\begin{equation}
A_1(\epsilon) = \epsilon^{D-2} \, e^{\epsilon\mu(\epsilon)}\;\; ;\;\; 
B_1(\epsilon) = \epsilon^{-2}\, \left\lbrace e^{\epsilon \mu(\epsilon)} 
- 2D\right\rbrace. 
\end{equation}
Similarly, ${\cal G}_c$ becomes, in the same limit
\begin{equation}
{\cal G}_c(\x) = \int_0^\infty dq\, J_1(q) {\cal H} (q,\x) 
\end{equation}
with
\begin{equation}
{\cal H} (q,\x) = \int { d^D {\bf p} \over (2\pi)^D} 
{A_2(\epsilon,q) e^{-i{\bf p}\cdot \x}\over p^2 + B_2(\epsilon,q)},
\end{equation}
where
\begin{eqnarray}
A_2(\epsilon,q) &=& - \epsilon^{D-2} \, e^{\epsilon\left[ \mu(\epsilon) 
+ (q^2/4\lambda(\epsilon))\right]}\\
B_2(\epsilon,q) &=& \epsilon^{-2}\, \left\{e^{\epsilon 
\left[ \mu(\epsilon) + q^2/4 \lambda(\epsilon)\right] } - 2D\right\}. 
\end{eqnarray}
The continuum propagator is defined  as
\begin{equation}
G(\x) = \lim_{\epsilon\to 0} \left\{ {\cal M} (\epsilon) 
{\cal G}(\x; \epsilon)\right\}, 
\end{equation}
where the small $\epsilon$ behavior of ${\cal M}(\epsilon),\lambda 
(\epsilon)$ and $\mu(\epsilon)$ have to be fixed in such a manner that 
this limit is finite.
One can easily see that, finiteness of $A_1$ and $B_1$ requires
\begin{eqnarray}
\lim_{\epsilon\to 0}\left\{ {\cal M}(\epsilon) \epsilon^{D-2} 
e^{\epsilon \mu(\epsilon)} \right\}&=& 1\\
\lim_{\epsilon\to 0} \left\{ {1\over \epsilon^2} 
\left[ e^{\epsilon \mu(\epsilon)} - 2D \right] \right\} &=& m^2, 
\end{eqnarray}
which is a standard result leading to~(\ref{eqn:contori}) 
(see {\it e.g.},~\cite{paddy94}). 
The finiteness of $B_2$ requires the quantity $\beta(\epsilon) \equiv 
\left[ \epsilon/ \lambda(\epsilon)\right]$ to scale as $\epsilon^2$ for 
small $\epsilon$. 
Writing $\beta(\epsilon) \simeq l_0^{-2} \epsilon^2 + {\cal O} (\epsilon^3)$ 
in this limit (where we expect $l_0\propto L_P$ in the continuum limit), 
we find that the final result can be expressed as $G = G_0 + G_c$, with
\begin{eqnarray}
G_0(\x) &=& \int {d^D {\bf p} \over (2\pi )^D} {e^{-i{\bf p}\cdot \x} 
\over p^2 + m^2},\\
G_c(\x) &=& - \int_0^\infty dq J_1(q) \int {d^D {\bf p} \over (2\pi )^D}\;
 e^{-i{\bf p}\cdot \x}\nonumber\\ 
& &\qquad\qquad\quad
\times\, \left\lbrace{1 \over p^2 + (D/2l_0^2)q^2 +  m^2}\right\rbrace.
\end{eqnarray}
Integrating the second term by parts and combining with the first term, 
we can express the full momentum space propagator as
\begin{equation}
\hat G({\bf p}) = 2 \nu^2 \int_0^\infty dq \ {q J_0 (q) \over 
\left[ q^2 +  \nu^2 (p^2 + m^2) \right]^2}, 
\end{equation}
where $\nu^2\equiv (2l_0^2/D)$. 
Using the identity
\begin{equation}
\int_0^\infty dx {xJ_0(x) \over (x^2 + b^2)^2} 
= - {1\over 2b} K_0^\prime (b) = {K_1(b) \over 2b}, 
\end{equation}
where $K_{n}(z)$ is the modified Bessel function of order $n$, we can write
\begin{equation}
\hat G ({\bf p}) = {\nu\over \sqrt{p^2 + m^2} } 
K_1 \left(\nu\sqrt{p^2 + m^2} \right).\label{eqn:finres}
\end{equation}

This is our final result with $\nu\propto L_P$ in the continuum limit. 
This equation represents the Feynman propagator for a ``free'' spin-zero 
particle when our prescription ---that the weightage for a path of length
$l$ should be invariant under the transformation $l \to L_P^2/l$---has 
been invoked. 
This postulate (which in the present context may be called `lattice duality')
and the form of the standard free particle propagator uniquely leads to our
final result. 
From the asymptotic forms of $K_1(z)$ it is easy to see that the propagator 
in~(\ref{eqn:finres}) has the limiting expressions 
\begin{equation}
\hat G({\bf p}) \rightarrow \cases{ {1\over p^2+m^2} & (for $\nu
\sqrt{p^2+m^2} \ll 1$)\cr
\exp(- \nu \sqrt{p^2+m^2}) & (for $\nu \sqrt{p^2+m^2} \gg 1$).\cr}
\end{equation}
When $ \nu \propto L_P \to 0$, the propagator reduces to the standard form 
while for energies larger than Planck energies it is exponentially damped. 

I shall now show that the result in~(\ref{eqn:finres}) has an extremely 
simple interpretation and an alternative derivation. 
The standard Feynman propagator in equation~(\ref{eqn:contori}) can be
equivalently represented as 
\begin{eqnarray}
G_{\rm conv} (\x) &=& \int {d^D p \over (2\pi )^D} 
{e^{-i{\bf p}\cdot \x } \over p^2+m^2}\nonumber\\ 
&=& \int {d^D p \over (2\pi )^D} e^{-i{\bf p}\cdot \x} 
\int_0^\infty ds \ e^{-s(m^2+p^2)} \nonumber\\
&=& \int_0^\infty {ds\over (4\pi s)^{D/2}} \exp \left(- {x^2\over 4s} 
- m^2s \right).
\end{eqnarray}
The last expression, in fact, constitutes the Schwinger's proper time 
version of the propagator. 
Suppose we  now postulate that the net effect of quantum fluctuations is 
to add a ``zero-point length'' to spacetime interval; ie., to  change 
the interval from $(x-y)^2$ to $(x-y)^2 + l_0^2$ where $l_0 \propto L_P$. 
(In~\cite{paddy87,paddy85}, it was suggested that $l_0 = L_P/2\pi$.) 
Making this replacement and doing the inverse Fourier transform, we 
immediately see that the modified momentum space propagator becomes 
\begin{equation}
\hat G_{\rm mod} ({\bf p}) = {l_0\over \sqrt{p^2+m^2}} K_1 
\left( l_0 \sqrt{p^2+m^2} \right),
\end{equation}
which is identical in form to equation~(\ref{eqn:finres}). 
In other words, {\it the modification of the path integral based on the 
principle of duality leads to results which are identical to adding a
``zero-point length'' in the spacetime interval}. 

I wish to argue that the connection shown above is non-trivial; 
I know of no simple way of guessing this result. 
The standard Feynman propagator of quantum field theory can be obtained 
either through a lattice regularization of a path integral or from 
Schwinger's proper time representation. By adding a zero-point 
length in the Schwinger's representation we obtain a modified propagator. 
 Alternatively, using the principle of duality, we could modify the 
expression for the path integral amplitude on the lattice and obtain---in 
the continuum limit---a modified propagator. 
Both these constructions are designed to  suppress energies larger than 
Planck energies. 
{\it However, there is absolutely no reason for these two expressions 
to be identical.} 
The fact that they are identical suggests that the principle of duality 
is connected in some deep manner with the spacetime intervals having a 
zero-point length.  
Alternatively, one may conjecture that any approach which introduces a 
minimum length scale in spacetime (like in string models) will lead to 
some kind of principle of duality. 
This conjecture seems to be true in conventional string 
theories~\cite{ashoke94,giveon94} though it must be noted that the term 
duality is used in somewhat different manner in string theories. 

I stress that the path integral amplitude is modified on the lattice 
{\it before} taking the continuum limit. 
This allows us to introduce a factor $\exp(-\lambda/N\epsilon)$ along with 
the original $\exp(-\mu\epsilon N)$. Loosely speaking, we are changing the
infinitesimal action for the relativistic particle from $ds$ to 
$(ds + L^2/ds)$. 
It is not easy to interpret this term directly in the continuum limit or 
even find a  modified {\it continuum} action for the relativistic particle 
which will lead to the same final propagator. 
It would be interesting to see whether this could be done. 
This question and related issues are under investigation.

I have been discussing these ideas with many people over the years.
In particular, I thank K.~Subramanian, L.~Sriramkumar and K.~Srinivasan 
for several useful discussions.


\begin{thebibliography}{20}
\bibitem{paddy87}
T.~Padmanabhan, Class.\ Quantum Grav.\ {\bf 4}, L107 (1987).  
\bibitem{paddy85}
T.~Padmanabhan, Ann.\ Phys.\ (N.Y.)\ {\bf 165}, 38 (1985). 
\bibitem{rovelli95}
C.~Rovelli and L.~Smolin, Nucl.\ Phys.\ B\ {\bf 442}, 593 (1995). 
\bibitem{ashtekar92}
A.~Ashtekar, C.~Rovelli and L.~Smolin, Phys.\ Rev.\ Lett.\ {\bf 69}, 
237 (1992). 
\bibitem{paddy94}
T.~Padmanabhan, Found.\ Phys.\ {\bf 24}, 1543 (1994).  
\bibitem{paddy96}
T.~Padmanabhan, Paper in preparation.
\bibitem{ashoke94}
A.~Sen, Int.\ J.\ Mod.\ Phys.\ A {\bf 9}, 3707 (1994).
\bibitem{giveon94}
T.~A.~Giveon, M.~Porrati and E.~Rabinovic, Phys.\ Rep.\ {\bf 244}, 77 (1994). 
\end{thebibliography}
\end{document}